\documentclass{elsart3p}

\usepackage{graphicx}

\usepackage{amsmath}
\usepackage{amssymb}
\usepackage{setspace}

\begin{document}

\begin{frontmatter}



\title{A generation-based particle-hole density-matrix renormalization group study of interacting quantum dots}


\author{Yuval Weiss\corauthref{cor1}}
\ead{weissy@mail.biu.ac.il}
\corauth[cor1]{Corresponding author}
\author{Richard Berkovits}
\address{The Minerva Center, Department of Physics, Bar-Ilan University, Ramat-Gan 52900, Israel}

\begin{abstract}
The particle-hole version of the density-matrix renormalization-group method (PH-DMRG) is utilized
to calculate the ground-state energy of an interacting two-dimensional quantum dot.
We show that a modification of the method, termed generation-based PH-DMRG, 
leads to significant improvement of the results, and
discuss its feasibility for the treatment of large systems.
As another application we calculate the addition spectrum.
\end{abstract}

\end{frontmatter}


\section{Introduction}
Recently there is a growing interest in the low-temperature physics
of disordered many-particle systems, such as electron dephasing due to interactions
\cite{aleiner99} and the two-dimensional (2D) 'metal-insulator' transition \cite{abrahams01}. 
Transport properties through quantum dots (QDs) have also been recently investigated and shown 
to exhibit interesting behavior in the presence of both interactions
and disorder \cite{alhassid00}.
An analytical treatment of these problems is unfortunately
difficult, since both the disorder and the interactions cannot be
considered as a small perturbation. Exact numerical methods for these problems are
restricted to small systems, since the size of the many-particle Hilbert space
grows exponentially with the system size.

During recent years, several methods were used to 
decrease the Hilbert space to a size which is computationally feasible. 
One way is to define an iterative order in which the system is treated, and use a 
smart truncation method between the iterations to reduce the space size. This is the 
idea behind the ensemble of renormalization group methods, such as the numerical
renormalization group (NRG) \cite{wilson} and the density-matrix renormalization 
group (DMRG) \cite{dmrg}.

A different approach uses a predefined constraint in order to truncate.
Thus, the entire system may be treated at once, yet not all the system states 
are taken into account. Therefore, the matrix size one needs to diagonalize is smaller 
than the entire Hilbert space dimension, and hopefully small enough to be exactly solved.
For example, one can approximate the ground state (GS) energy of a system by considering only part of 
the single-level eigenstates. A sorted list of the eigenstates can be obtained by performing 
a self-consistent Hartree-Fock (HF) calculation, after which the eigenvectors with the highest
eigenvalues are neglected \cite{berkovits03-2}. 
Such states should, intuitively, have the smallest contribution to the many-particle GS energy. 

Another suggestion is based on the localization of the Fock space. 
Since the interaction term is a two-body 
operator, only many-body states which differ by at most two electron-hole pairs are coupled by 
the Hamiltonian. It was shown \cite{altshuler97} that the average contribution of a state 
containing $k$ electron-hole pairs to the exact GS is proportional to $\exp(-k/{\xi_F})$, 
where $\xi_F$ is the Fock space localization length. Considering also the number of
states in the $k$-th electron-hole generation, $\binom{N-n_e}{k} \binom{n_e}{k}$, 
where $n_e$ is the number of electrons and $N$ is the number of single-particle states,
one finds \cite{berkovits03-2} that the weight of generations falls off exponentially 
as long as $\xi_F^{-1} > \ln[(N-n_e)n_e/(k+1)^2]$.
Thus one can consider in the approximated Hamiltonian only states with a small 
number of particle-hole (PH) pairs.  

In the following we denote these global truncation methods as energy-truncation (ET) and generation-truncation (GT).
For a full discussion of both see Ref.~\cite{berkovits03-2}.

Unfortunately, one cannot simply implement these global truncation methods for larger systems. 
As the size of the system increases, the number of energy states or PH generations, 
for a given cutoff, increases, and thus the matrix size increases exponentially, and it soon becomes too 
large to be solved exactly. Therefore, an alternative method is needed, which can give an accurate
approximation to the GS energy, yet can be extended to larger systems.
In the following we show that a generation-based PH-DMRG method is suitable for that task.

\section{Model}
\label{sec:ph_model}
We model the QD as a 2D disordered lattice occupied by
$n_e$ spinless electrons. Hopping between nearest neighbors (NNs),
and either NN or Coulomb interactions are considered. Therefore, the Hamiltonian 
can be written as $\hat H = \hat H_0 + \hat H_{\rm int}$, where 
\begin{eqnarray} \label{eqn:H_phdmrg}
{\hat H}_0 &=& \sum_{m} \epsilon_{m}{\hat a}^{\dagger}_{m}{\hat a}_{m} 
-t \displaystyle \sum_{\langle m,n \rangle}({\hat a}^{\dagger}_{m}{\hat a}_{n} + H.c.),
\end{eqnarray}
and $\hat H_{\rm int}$ is one of the terms
\begin{eqnarray} \label{eqn:H_int_nn}
{\hat H}_{\rm int}^{\rm (NN)} &=& 
\displaystyle V \sum_{\langle m,n \rangle}{\hat a}^{\dagger}_{m}{\hat a}^\dagger_{n}{\hat a}_{n}{\hat a}_{m}; \\ \nonumber
{\hat H}_{\rm int}^{\rm(C)} &=& 
\displaystyle \frac{1}{2} \sum_{m \ne n}\frac{V_c}{|r_m-r_n|}{\hat a}^{\dagger}_{m}{\hat a}^{\dagger}_{n}{\hat a}_{n}{\hat a}_{m}.
\end{eqnarray}
Here, $\hat a^\dagger_m$ and $\hat a_m$ denote creation and annihilation operators
of an electron in lattice site $m$, $\langle \cdots \rangle$ represents NNs,
$\epsilon_{m}$ is the on-site energy of site $m$, chosen randomly from a uniform distribution
$[-W/2,W/2]$, and $t$ is the hopping matrix element between NNs which is conventionally taken as the energy unit.
In the interaction terms, $V$ is the NN interaction strength, $V_c$ is the Coulomb interaction in a distance of one
lattice unit, and $|r_m-r_n|$ is the distance between sites $m$ and $n$, measured in lattice units.

The GS energy of the system is calculated using the PH-DMRG 
method \cite{dukelsky}. A detailed description of the specific
implementation can be found elsewhere \cite{dissertation}.
The general idea of the method is to divide the HF single-particle energy levels to those above 
(particle-states) and below (hole-states) $E_F$, and then treat these states iteratively. 
Starting from the vacuum state, in which all levels below $E_F$ are filled, 
and all the others are empty, in each iteration one particle-state and one
hole-state are added, starting from $E_F$ and proceeding in both directions. 
The superblock is composed of the states already added, by maintaining the number of particles constant. 
The superblock diagonalization is followed by a truncation of the Hilbert space of both the particle 
block and the hole block, using their corresponding density matrices. The iteration ends, and a new 
couple of states can be added again. The process stops after all of the states were added.

The accuracy of the PH-DMRG method depends 
mostly on the number of states, $p$, that are kept between successive iterations. In other words,
in each iteration, $p$ eigenvectors of the density matrix are taken, while all the others are 
neglected. Except the accuracy, $p$ influences dramatically the computational resources 
required during the process. 

In order to compare the PH-DMRG method to other approximations we define
the error rate, or the discrepancy, as $D(x) = |x^\prime-x|/|x|$,
where $x$ is an exact quantity and $x^\prime$ is an approximation for $x$.
By calculating $\bar D(E_{GS})$, averaging over different realizations 
of the disorder, one thus has a good estimate about the accuracy of the approximation method.

\section{Ground-State Energy Calculation}
\label{sec:ph_res}
In order to examine the accuracy of the PH-DMRG method we investigate the case of 
a $4 \times 6$ lattice with a disorder strength $W=5t$, with the
same disorder ensemble used in Ref.~\cite{berkovits03-2}. In this section we restrict ourselves to 
the case of NN interactions, with strength $V=3t$, and use $n_e=10$.

\begin{figure}[htb]\centering
\vskip 0.5truecm
\begin{minipage}[t]{.45\textwidth}
\includegraphics[width=.95\textwidth]{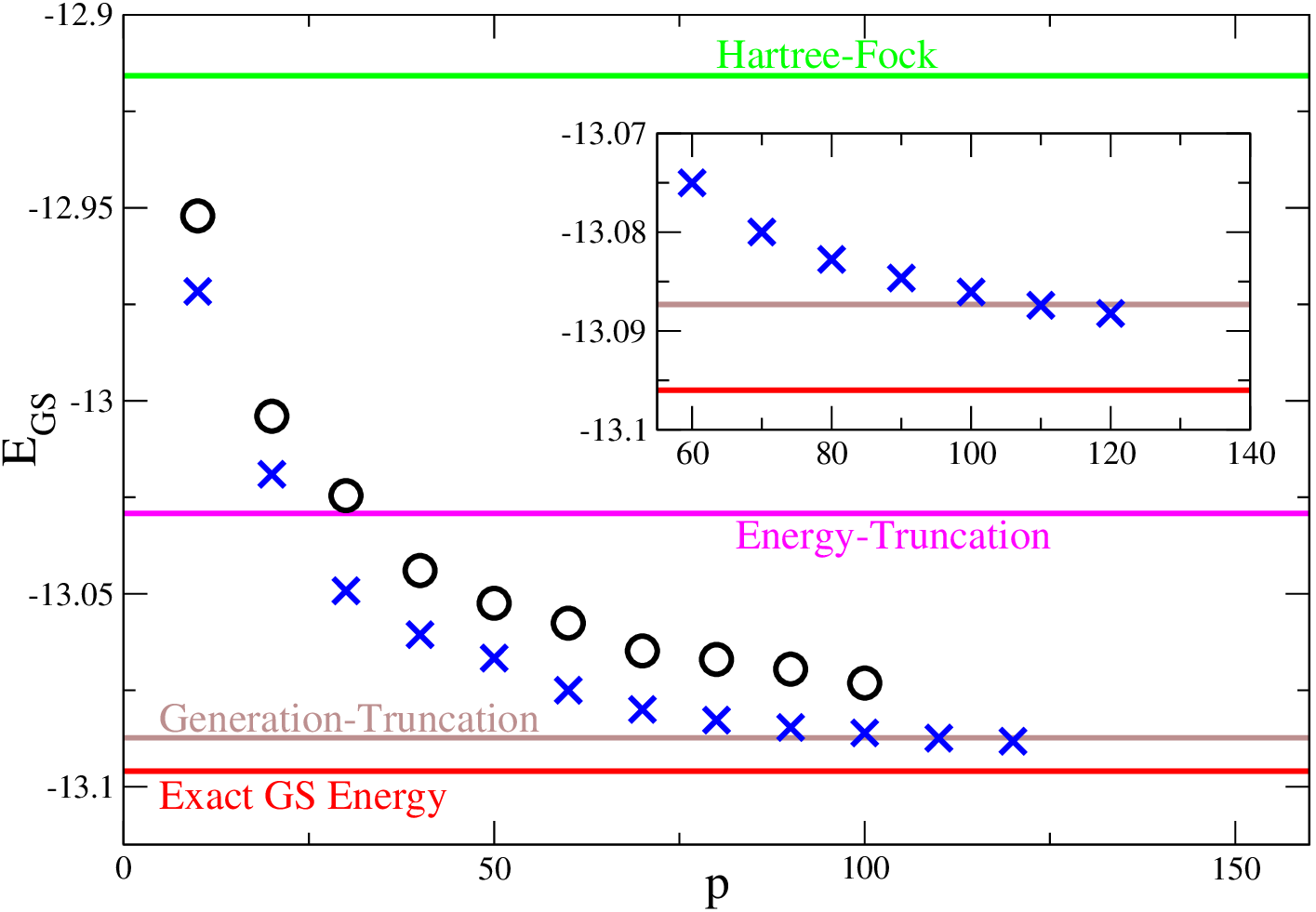}
\end{minipage}
\hfil
\vskip 1truecm
\begin{minipage}[t]{.45\textwidth}
\includegraphics[width=.95\textwidth]{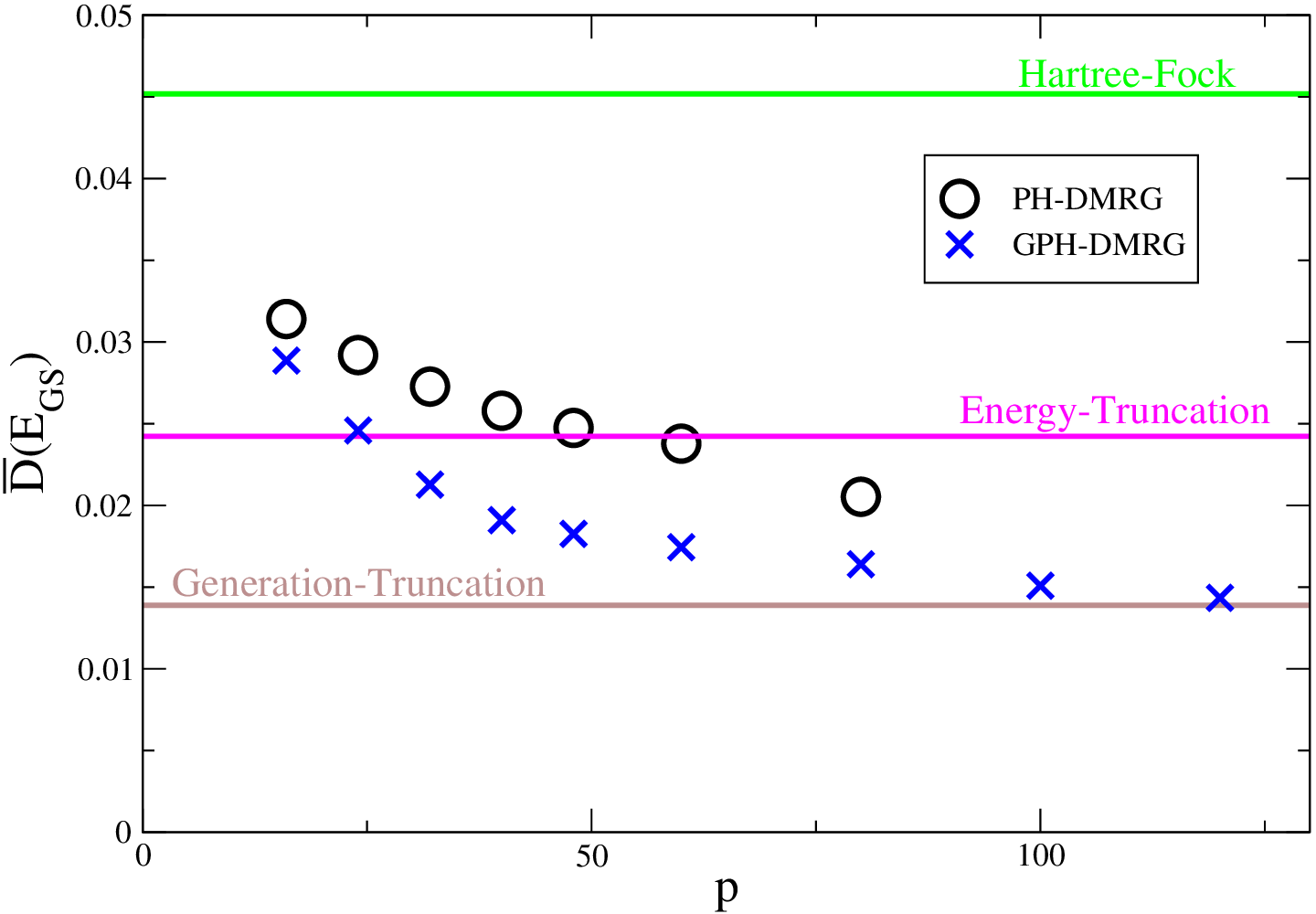}
\end{minipage}
\caption[Energy calculation and discrepancy using PH-DMRG]
{\label{phfig_1}
PH-DMRG results for the GS energy calculation as a function of $p$
(regular version - circles, generation-based version - crosses),
compared to various approximated results (see text).
(a) The GS energy of a specific realization.
Inset: zoom into the region of $p \ge 60$. 
(b) The discrepancy $\bar D(E_{GS})=\langle|E_{GS}'-E_{GS}|/|E_{GS}|\rangle$ averaged
over an ensemble of $100$ realizations. 
}
\end{figure}

Typical results, for a specific realization, are shown in Fig.~\ref{phfig_1}(a) (circles),
as a function of $p$, the number of block states kept. 
The results are compared to the HF results, to the results of the ET method, 
in which the lowest $18$ HF states were used, and to those of the GT method, 
in which up to $3$ PH generations were kept. The exact results are also drawn.
As $p$ increases, the PH-DMRG approximation improves. 

Averaging over realizations (Fig.~\ref{phfig_1}(b) (circles)) shows that
the average accuracy of the PH-DMRG calculation improves very slowly by increasing $p$.
The best PH-DMRG results shown (for keeping $80$ states) gets closer to the exact GS
than the ET method, yet the GT method is much more accurate.

The comparison between the PH-DMRG results and those of the GT method raise the
following question:
Can one improve the PH-DMRG process without
increasing $p$? In the following we present such an
improvement, motivated by an analysis of the PH-DMRG truncation method.
In essence, the main difference between the DMRG and NRG methods,
is the truncation algorithm. In the NRG method, the truncation of states is based on their 
energies, while in the DMRG it is based on the density-matrix eigenvalues.
The density-matrix eigenvectors with the highest eigenvalues are considered as the most 
important, and the rest of the states are neglected.

Nevertheless, this difference is not the main reason for the DMRG success.
This success originates from the fact that while the NRG truncation 
method is based only on the sites which were already iteratively added,
the DMRG algorithm is influenced by additional sites included in the superblock.
The superblock is composed of a "system" coupled to an "environment",
which represents all the sites which were not yet included in the iteration process,
and thus the truncation is based on extrapolation, which leads to better results.

On the other hand, the superblock in the PH-DMRG does not consider any "future" states.
The PH-DMRG process couples only the current particle-states and hole-states, but
other states
are not part of the superblock. 
Therefore, the truncation does not take them into account, and this limits its success.

In order to improve the truncation decision we suggest to improve the PH-DMRG method 
by taking into account the number of PH generations as an extra 
condition taken during the truncation step. 
As we have discussed above, the idea of the Fock space localization indicates that
the weight of successive PH generations decreases exponentially.
Therefore, one expects that among states belonging to a given iteration,
those with a smaller number of PH generations are more important.

In order to incorporate this criterion into the density-matrix truncation process,
we retain all states which contain less than $k$ PH generations in the first round.
If there are too many such states, we use the density-matrix eigenvalues, and take the states
with the highest eigenvalues. After this first round, if there is still a room for more 
states, we retain also states with $f > k$, according to their density-matrix eigenvalues,
until the maximal number of states $p$ is reached.

For the $4 \times 6$ system with $10$ electrons, we've executed the suggested
truncation method for $k=0,1,2,3,4$, and for each realization picked the
lowest energy. It turns out, however, that in $95$ percents of the samples the $k=2$ case 
is the most accurate, so that using a constant $k=2$ would lead to similar results.
These results are also shown in Fig.~\ref{phfig_1} (crosses).
As can be clearly seen, the results obtained by the improved method, 
termed generation-based PH-DMRG (GPH-DMRG), are better 
by almost $30$ percents than the regular PH-DMRG results, and are comparable to the 
results of the GT method of Ref.~\cite{berkovits03-2}.

\section{Long-range interactions; Strong interactions}
\label{sec:ph_cnn}

The results of the previous section were obtained for short-range interactions. 
In the framework of the real-space DMRG method 
there is a huge impact when the interaction range increases. The DMRG iterations add
subsequent sites one after another, and if long-range interactions are considered, much more 
data should be stored from previous steps. Practically, therefore, real-space DMRG applications 
traditionally consider only short-range interactions.

On the other hand, the interactions range does not affect the GPH-DMRG method almost at all.
It does of course change the self consistent HF stage, however,
the entire iterations process, which is the most taxing stage of the 
GPH-DMRG algorithm, remains the same. Thus, one may
ask what effect does the interactions range have on the GPH-DMRG accuracy. 
A related question is how the accuracy is changed when the interactions become stronger.

\begin{figure}[h]\centering
\includegraphics[width=7cm,height=!]{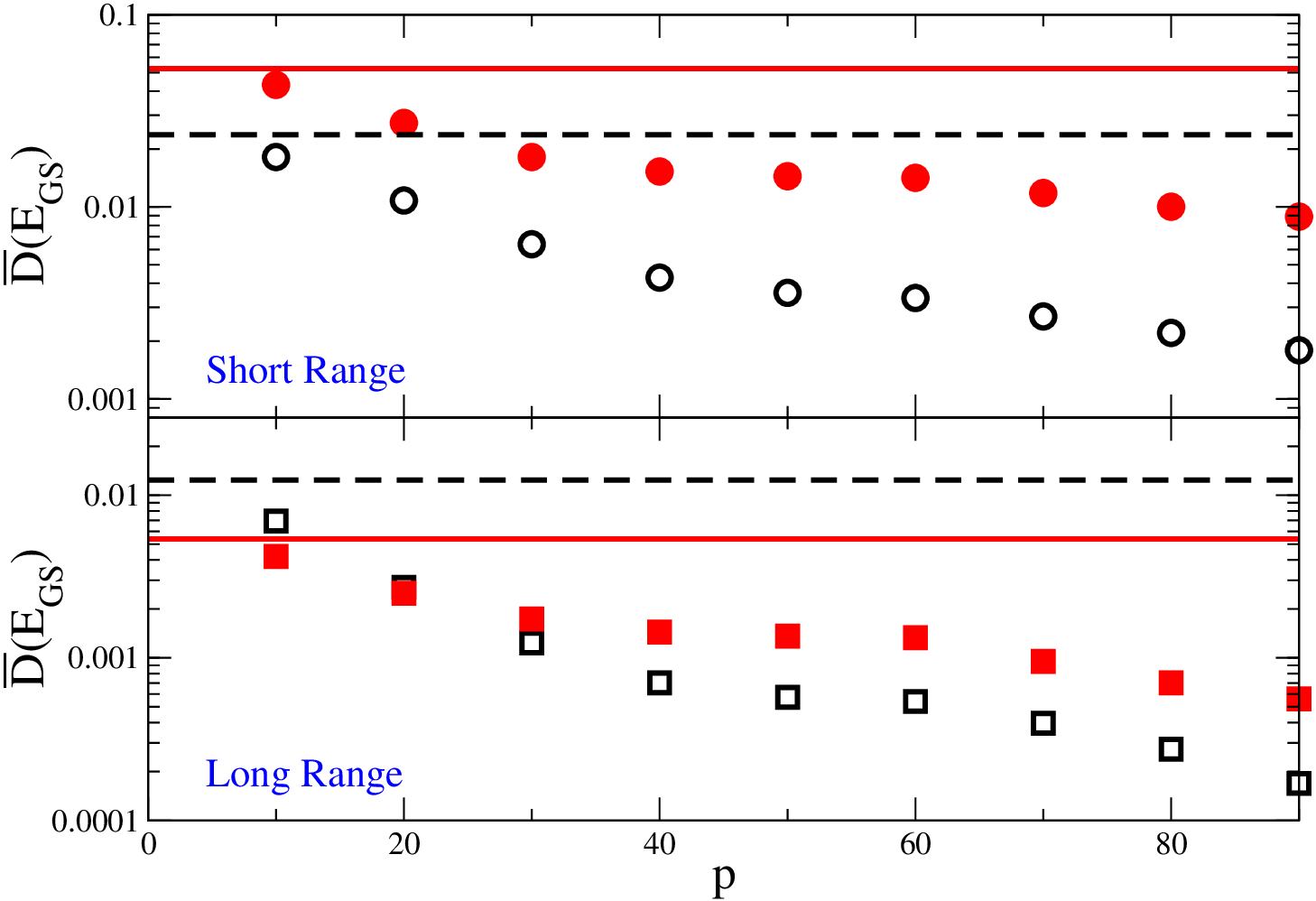}
\caption[GPH-DMRG results for short-range and long-range interactions]
{\label{phfig_1c}
The discrepancy $\bar D(E_{GS})$ obtained by the GPH-DMRG calculation as a function of the number of states kept,
for an ensemble of $4 \times 6$ disordered lattices occupied by $6$ electrons. The accuracy is compared between 
short-range (upper panel) and long-range (lower panel) interactions, and between 
intermediate interactions strength ($V=V_c=3t$, empty symbols) and a strong one 
($V=V_c=10t$, filled symbols).
The solid (dashed) lines show the results of the HF approximation for strong (intermediate) interactions.
Note the semi-log scale.
}
\end{figure}

In Fig.~\ref{phfig_1c} we present the discrepancy between the exact GS and the one 
obtained by the GPH-DMRG for short-range (upper panel)
and long-range (lower panel) interactions. Both strong (filled symbols) and 
intermediate (empty symbols) interaction strengths are considered.
Since these results were obtained for lattices with $6$ electrons, i.e., 
the number of hole-states is small, the accuracy of both the HF approximation and the
GPH-DMRG method is better than in the previous section. However,
the results show that for both interaction types the GPH-DMRG accuracy is reduced when the 
interaction strength is enhanced.

A striking feature is the difference in the discrepancy between the short-range and the long-range 
interactions. For small values of $p$, the improvement in accuracy of the 
long-range case is explained by the fact that the HF approximation is
better. Nevertheless, the improvement of the long-range results by the 
GPH-DMRG method is fascinating ($\bar D$ for $V_c=3t$ is reduced by two orders of magnitude), 
and the averaged discrepancy (for $V_c=3t$ with $p=90$ states) is 
$\sim 10^{-4}$, more than an order of magnitude better than the accuracy in the 
short-range case. 

\section{Addition Spectrum Calculation}
\label{sec:ph_delta2}
As a useful application of the GPH-DMRG we calculate
the addition spectrum of a QD, and compare the discrepancy to that of the HF approximation.
The addition spectrum can be defined by 
\begin{eqnarray} \label{eqn:ph_Delta2_exact}
\Delta_2 = E_{GS}(n_e) - 2E_{GS}(n_e-1) + E_{GS}(n_e-2).
\end{eqnarray}
Therefore, for a calculation of $\Delta_2$ one needs the GS
energies of $3$ successive electron numbers for each realization, and we choose 
the ensemble of $4 \times 6$ samples used in the previous sections, 
occupied by $4,5$ and $6$ electrons, with NN or Coulomb interactions. 
In general, the results are
better in the Coulombic case, because of a higher accuracy for each energy calculation. 

\begin{figure}[h]\centering
\vskip 1truecm
\includegraphics[width=6cm,height=!]{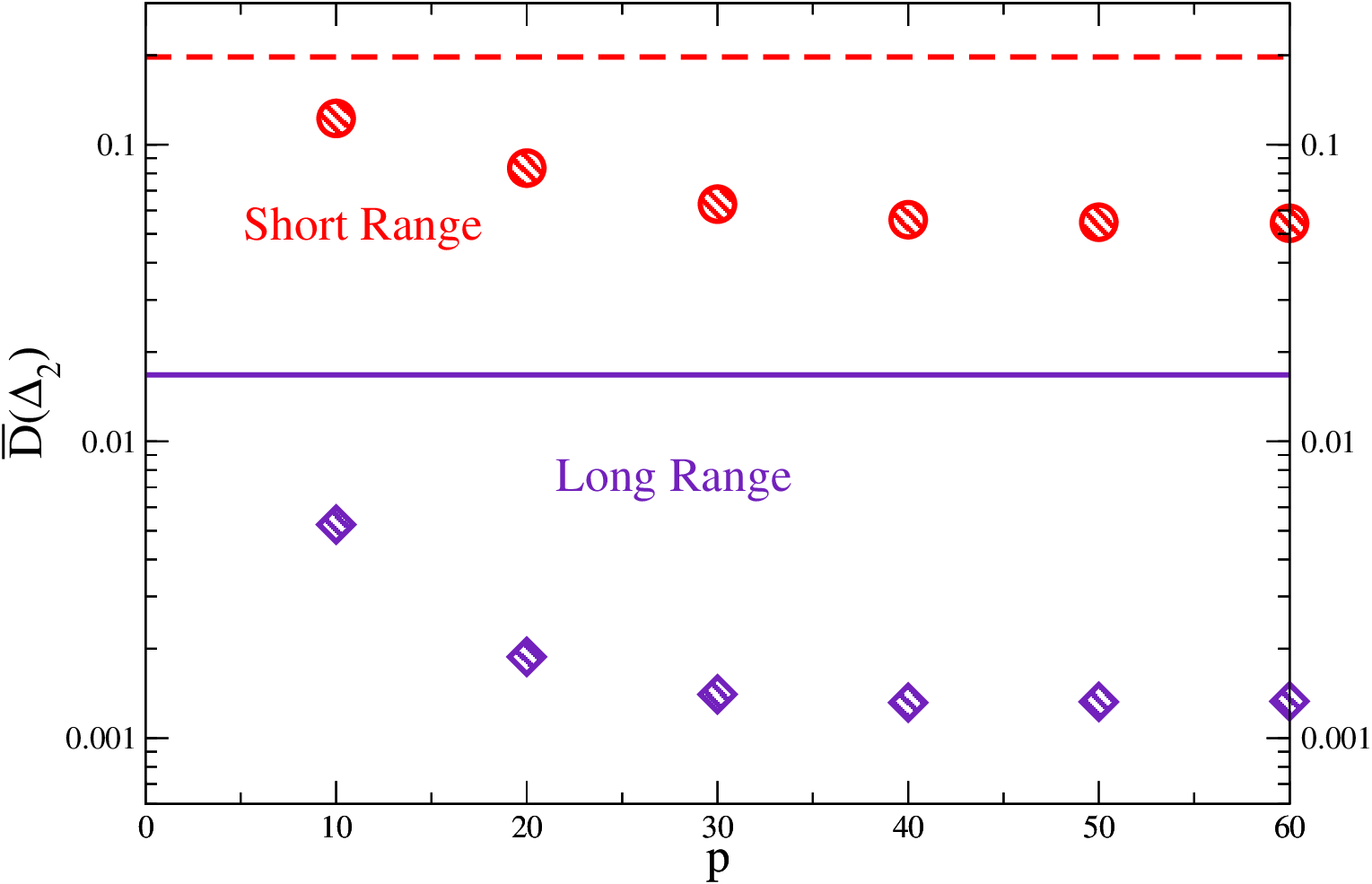}
\caption[Discrepancy of $\Delta_2$ calculation using GPH-DMRG]
{\label{phfig_2}
The averaged discrepancy $\bar D(\Delta_2)$ obtained by the  
GPH-DMRG calculation, for NN (dashed line and circles) and Coulomb (solid line and diamonds) interactions. 
The GPH-DMRG results are shown in a semi-log scale as a function of $p$ (symbols), together with the HF results (lines).
}
\end{figure}

In Fig.~\ref{phfig_2} the results for the averaged discrepancy of $\Delta_2$ are presented.
The HF approximation obtains, for Coulomb interactions, $\bar D(\Delta_2) \approx 1.7$ percents. 
The GPH-DMRG method, even with a very small number of states ($p \ge 30$), 
exhibits an improvement of the error rate by more than an order of magnitude,
to a level of $\sim 0.13$ percents.

\begin{figure}[h]\centering
\vskip 0.7truecm
\includegraphics[width=7cm,height=!]{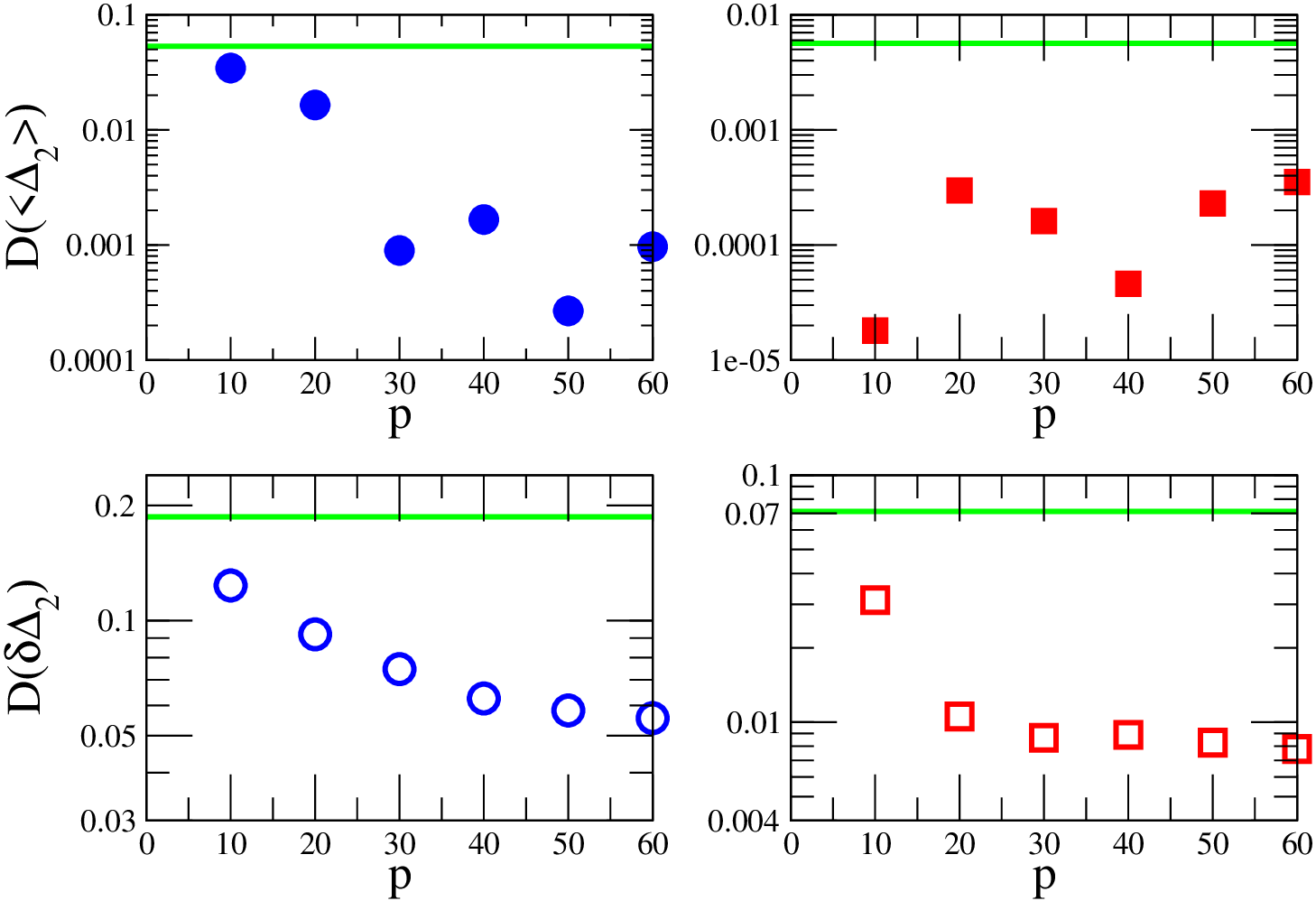}
\caption[Calculation of $\langle \Delta_2 \rangle$ and $\delta \Delta_2$ with short-range and long-range interactions]
{\label{phfig_3}
The GPH-DMRG results for the discrepancy of $\langle \Delta_2 \rangle$ (filled symbols) and
$\delta \Delta_2$ (empty symbols) of the $4 \times 6$ systems occupied by $4,5$ and $6$ 
electrons with either NN (circles) or Coulomb (squares) interactions.
The lines correspond to the HF results.
}
\end{figure}

However, in the NN case the results are quite poor. The starting point of the GPH-DMRG algorithm, i.e.,
the HF results, give an average error of almost $20$ percents. The GPH-DMRG improves it by a factor of $4$, 
to the order of $5$ percent, which is still a very high error rate.
It is thus interesting to check whether the calculation of $\langle \Delta_2 \rangle$ and 
$\delta \Delta_2 = \sqrt{ \overline{ \left( \Delta_2- \overline{\Delta_2} \right)^2}}$ are more accurate.
These results are shown in Fig.~\ref{phfig_3}.

As can be seen, the results for $\langle \Delta_2 \rangle$ are very accurate for both 
interaction types, even for very small $p$. Notice that the results
continue to fluctuate around the exact result, since the approximation
for $\Delta_2$ is done using $3$ different GS approximations, which have different
convergence rates. However, for $p \ge 30$, the error is less than $0.1$ percent in both cases.
The results of $\delta \Delta_2$, on the other hand, show slow convergence.
The error rates obtained by the GPH-DMRG (using $p=60$)
for the different cases are summarized in Table~\ref{tbl:phdmrg_res_d2}.
The ratio between the HF discrepancy and that of the GPH-DMRG is shown in parentheses.

\begin{table}[ht] 
\centering 
\begin{tabular}{|c||c|c|}
\hline
&{\bf NN}&{\bf Coulomb}\\ [1ex]
\hline \hline  
$\bar D(\Delta_2)$& 0.054314 (3.6) & 0.001330 (12.6) \\ [1ex]
$D(\langle \Delta_2 \rangle)$& 0.000965 (55.1) & 0.000352 (16.0) \\ [1ex]
$D(\delta \Delta_2)$& 0.055554 (3.4) & 0.007794 (9.2) \\ [1ex]
\hline
\end{tabular}
\vskip 0.3truecm
\caption{GPH-DMRG error rates for $\Delta_2$ calculation} 
\label{tbl:phdmrg_res_d2}
\end{table}

It is easy to see that for Coulomb interactions the GPH-DMRG improves all results 
related to the addition spectrum by an order of magnitude, and leads to error rates of
less than $1$ percent for all cases. However, for the NN interactions, a small error rate
and a significant improvement are seen only for $\langle \Delta_2 \rangle$,
while modest factors are obtained for $\bar D(\Delta_2)$ and $\delta \Delta_2$,
with error rates larger than $5$ percents.

\section{Discussion}
\label{sec:ph_end}
In the previous sections we have seen that with an improvement of the 
truncation algorithm, the GPH-DMRG method can be used for an accurate approximation of 
the GS energy in disordered systems with interactions. 
In each of the cases we have checked, the GPH-DMRG led to a significant improvement
of the HF results. 
A comparison to the two methods of Ref.~\cite{berkovits03-2} 
shows that the GPH-DMRG error rate is better than that of the ET method,
and is similar to that of the GT one.

Nevertheless, for a full comparison one must also consider the feasibility of these methods 
when larger systems are treated. To understand the difference 
between the methods, we compare the matrix sizes for the case considered in section \ref{sec:ph_res}.
The ET, based on $18$ energy levels, and the GT, with up to $3$ PH
generations, require the diagonalization of matrices of sizes $43,758$ and $47,916$, respectively. 
The largest superblock Hamiltonian in the GPH-DMRG process, for $p=120$, 
is of size $13,494$.

When a larger lattice is studied, and in order to get the same accuracy as for the small system,
the ET and GT methods must include more levels and generations. 
Even was it sufficient to take the same number of levels and generations as in the small case, 
the size of matrices would have grown exponentially with the lattice size. Therefore, 
these methods become infeasible even for modest lattice sizes. In the GPH-DMRG method, 
on the other hand, the size of the matrix depends on the block size, 
and not on the system size. Nevertheless,
for larger systems, the number of single-particle states is larger,
and the discrepancy of the GPH-DMRG is expected to increase, unless $p$ is increased.
Therefore, the matrix size being diagonalized is expected to increase in the GPH-DMRG method as well. 

\begin{figure}[h]\centering
\vskip 1truecm
\includegraphics[width=5cm,height=!]{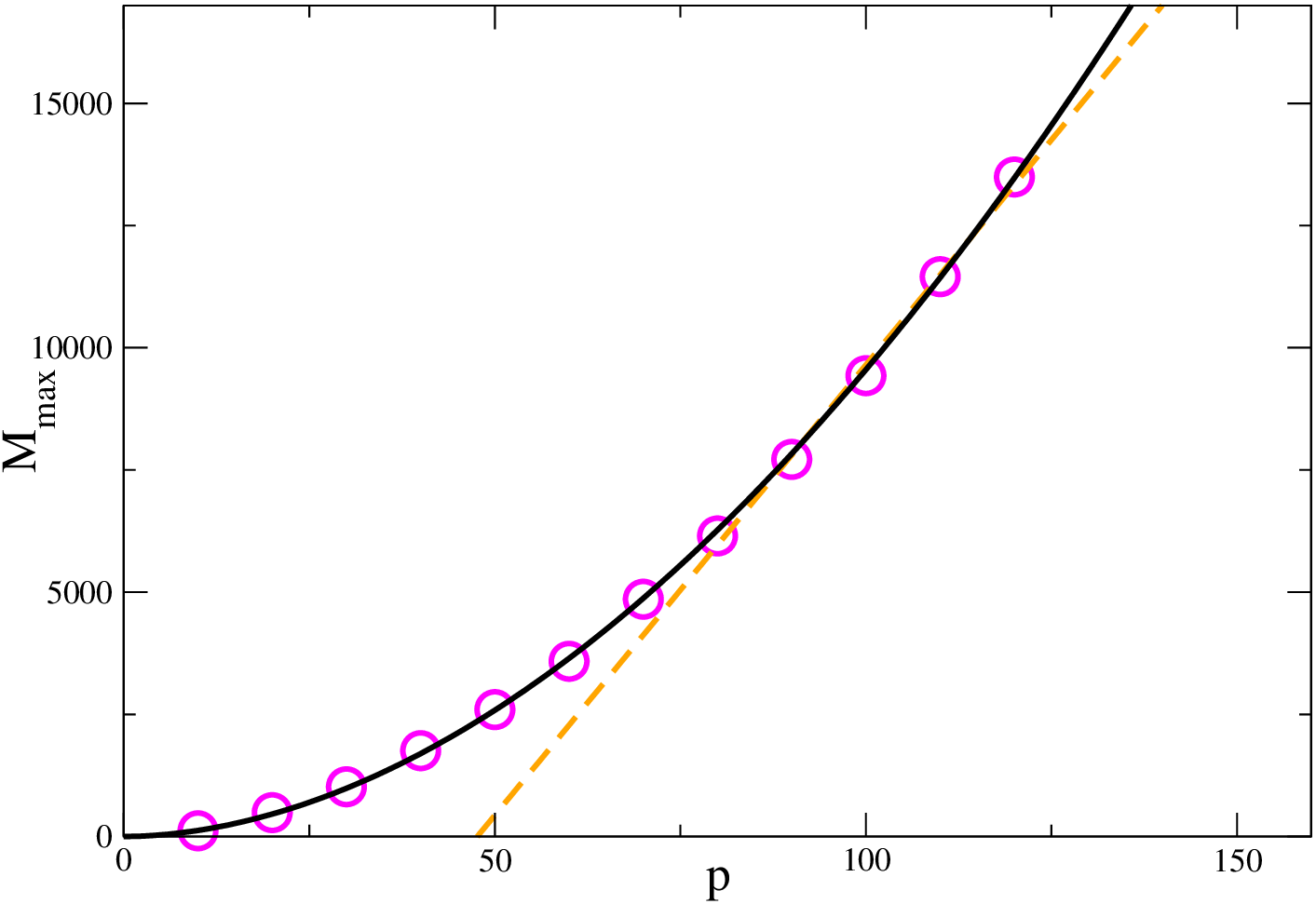}
\caption[Maximal superblock size]
{\label{phfig_6}
The maximal size of the superblock Hamiltonian needed to be diagonalized, for the
case of section \ref{sec:ph_res}. The dashed line is a linear fit of the
$p>80$ points, while the solid line is a power law fit in the entire range.
}
\end{figure}

The dependence of the largest matrix size needed to be diagonalized in the GPH-DMRG process, 
$M_{max}$, on $p$, is shown in Fig.~\ref{phfig_6}. Although for large values 
of $p$, a linear dependence was proposed \cite{dimitrova02}, from our results 
a power law $M_{max} \sim p^{1.89}$ seems more appropriate.
In any case, it is clear that the matrix size is less than quadratic in $p$.
Furthermore, since the largest matrix size used in our current GPH-DMRG application is still much 
below the technology limit, its increase should not be impossible.
It is thus clear that in principle the GPH-DMRG is capable of treating larger systems.

As a final remark we note that the addition spectrum of disordered QDs with interactions is 
a long-standing question in mesoscopic physics. Specifically, the behavior of $\delta \Delta_2$ 
for strong interactions is not completely understood. A comprehensive study of this question 
was performed a few years ago using the self-consistent HF method \cite{walker99}. Based on
our results, it may thus be relevant to re-examine that question using the GPH-DMRG method.

~

Support from the Israel Academy of Science
(Grant 877/04) is gratefully acknowledged.




\begin{thebibliography}{00}


\bibitem{aleiner99}
I. L. Aleiner, B. L. Altshuler and M. E. Gershenson, Waves Random Media {\bf 9}, 201 (1999).

\bibitem{abrahams01}
E. Abrahams, S. V. Kravchenko and M. P. Sarachik, Rev. Mod. Phys. {\bf 73}, 251 (2001).

\bibitem{alhassid00}
Y. Alhassid, Rev. Mod. Phys. {\bf 72}, 895 (2000).

\bibitem{wilson}
K. G. Wilson, Rev. Mod. Phys. {\bf 47} 773 (1975).

\bibitem{dmrg}
S. R. White, Phys. Rev. B {\bf 48}, 10345 (1993);
U. Schollw\"ock, Rev. Mod. Phys. {\bf 77}, 259 (2005).

\bibitem{berkovits03-2}
R. Berkovits, Solid State Commun. {\bf 127}, 725 (2003).

\bibitem{altshuler97}
B. L. Altshuler, Y. Gefen, A. Kamanev and L. S. Levitov, Phys. Rev. Lett. {\bf 78}, 2803 (1997).

\bibitem{dukelsky} 
J. Dukelsky, S. Pittel, S. S. dimitrova and M. V. Stoitsov, 
Phys. Rev. C {\bf 65}, 054319 (2002);
J. Dukelsky and S. Pittel, Rep. Prog. Phys. {\bf 67}, 513 (2004).

\bibitem{dissertation}
Y. Weiss, 
Ph.D. dissertation, Bar-Ilan University, Israel (2007).

\bibitem{dimitrova02} S. S. Dimitrova, S. Pittel, J. Dukelsky and M. V. Stoitsov,
in {\it The DMRG method for realistic large-scale nuclear shell-model calculations},
Proceedings of the 21st workshop on nuclear theory, edited by V. Nikolaev (Heron Press, 2002);
arXiv:nucl-th/0207025.

\bibitem{walker99}
P. N. Walker, G. Montambaux and Y. Gefen, Phys. Rev. B {\bf 60}, 2541 (1999).





\end{thebibliography}
\end{document}